# Spatial structure and aggregation of carbon allotrope nanofillers in isotactic polypropylene composites studied by small-angle neutron scattering

L.V. Elnikova[1], A.N. Ozerin[2], V.G. Shevchenko[2], P.M. Nedorezova[3], A.T. Ponomarenko[2], V.V. Skoi[4,5], A.I. Kuklin[4,5]

[1]NRC "Kurchatov Institute" – Alikhanov Institute for Theoretical and Experimental Physics, 117218 Moscow, Russia,
[2]Enikolopov Institute of Synthetic Polymeric Materials, RAS, 117393 Moscow, Russia,
[3]Semenov Institute of Chemical Physics, RAS, 119991 Moscow, Russia,
[4]Joint Institute for Nuclear Research, 141980 Dubna, Russia
[5]Moscow Institute of Physics and Technology, 141701 Dolgoprudny, Russia

**Abstract**

We study the aggregation of carbon allotrope nanofillers in the matrix of isotactic polypropylene with direct small-angle neutron scattering measurements. With the ATSAS software, we analyzed the data and determined the fractal shape, dimension, and sizes of nanofiller aggregation in the bulk of isotactic polypropylene over the range of the scattering angles. We estimated the volume distributions and aggregation of different types of carbon nanofillers at different concentrations: nanographite, graphene nanoplatelets (GNP), single-walled carbon nanotubes (SWCNT), multi-wall carbon nanotubes (MWCNT), binary fillers MWCNT/GNP and fullerenes. We reconstructed the shape of nanoscale aggregates of all nanofillers and found that the systems are polydisperse; nanofillers associate in the volume of iPP as fractal dense aggregates with rugged surface, their sizes exceeding original dimensions of nanofillers several times.

1. **Introduction**

The polymers filled with carbon allotropes have wide applications in different fields of human activity: industry, medicine, electronics and spintronics, sensing materials, lubricants, they are used as elements of electronic devices and actuators etc.[1,2]. The morphology of the fillers in polymer matrix determines functional properties of such composites when carbon allotropes are added, e.g. electrical and mechanical properties can change [3], these changes are also manifested due to size effects. Doping of polymer with carbon nanofillers is accompanied by modification of surface, changes in polymer structure and interactions between nanofillers.

Physical, chemical and structure properties of filled polymer nanocomposites can be investigated with different techniques, e.g. with dielectric spectroscopy, atomic force microscopy (AFM), differential scanning calorimetry (DSC), scanning electron microscopy (SEM), Raman scattering, Small-Angle X-ray scattering (SAXS) [3,4,5] *etc*.

However, during sample preparation one gets, as a rule, cleavages and cuts of the materials, that damage and distort their bulk structure. These additional defects result in a loss or distorted information about the test specimen. Whereas the SANS method provides nondestructive structure analysis, with length scale from 1 Å, while explicitly characterizing the morphology of nanofillers in the bulk of the material [4].

There is extensive SANS data for polymer nanocomposites [6-10], though very little if any information exists about iPP nanocomposites, which have specific chemical composition and bulk configuration of carbon allotrope nanofillers.

Various incorporated inhomogeneities in a polymer matrix correlate with the changes of the SANS curves (peak position, shape, slope, *etc.*) as compared to basic polymer. For interpretation of the experimental SANS curves, many models for identification of structures are applicable in frames of well-developed non-linear least square fitting models [4,11-14] for the Porod and Guinier scattering laws. In the bulk of polymers, nanofillers may form multilevel fractal objects [7-10, 13]; such aggregation changes the slope of the Guinier functions at appropriate scales of scattering angles.

Also, modern methods of small-angle scattering curves processing are supplemented with calculating procedures [11, 14] that make it possible to visualize the shape of particles in the bulk, to identify various morphologies of single nanoparticles as well as their aggregation, to distinguish mono- or polydispersity of the system *etc*. To reconstruct the shape of an object in the bulk from scattering spectra, the regularization methods are usually employed [12]. For example, in [11] and references therein, the capabilities the ATSAS program are reported.

The goal of our SANS experiments is to characterize the nanoscale effects in iPP matrix induced by the presence of carbon allotrope nanofillers (nanographites, GNPs, fullerenes, CNTs and their combinations) and to describe morphology of the formed nanoobjects. We are also motivated in possibility of further complex studies of physical properties of these composites induced by carbon nanofiller addition; basing on these SANS data, we can examine the dielectric and structure changes of these materials with electron spectroscopy, positron annihilation spectroscopy and other techniques.

## 2. Materials and methods

The samples of isotactic polypropylene (iPP) are filled by *in situ* polymerization with GNP at concentrations of 0.7 and 1.8 wt%, nanographite at 1.5 and 3.6 wt%, SWCNT at 1.2, 2.6 and 8 wt%, MWCNT at 3.5 wt%, MWCNT/GNP at 0.48, 0.9, 1.16 and 3 wt% and fullerenes at 16.5 wt %(the volume ratio for binary filler MWCNT:GNP is 1:2).

The chemical formula of iPP is $(C_3H_6)_n$, its density is 0.9–0.91 g/cm$^3$, and the degree of crystallinity is 60%. Nanographite particles are in the form of plates with diameter 112.7 nm and thickness of 47.3 nm.

Graphene nanoplatelets (GNP) were produced by chemical or thermal reduction of graphite oxide (TRGO) [15].

Graphite oxide was produced using modified method of Hammers - oxidizing graphite by $KMnO_4$ in concentrated $H_2SO_4$ [16-17].

X-ray diffraction analysis of GNP and TRGO powders was made using ADP-1 diffractometer [18]. For GNP and TRGO, the values of crystallite size were calculated to be 1.127 and 1.003 nm, respectively. Accordingly, synthesized few-layer particles are estimated to contain 3-5 layers of graphene. The approximate dimensions of individual GNP particle is 100 nm × 100 nm × 1.127 nm.

Pristine CVD-grown MWCNTs (purity C95%, average diameter < 10 nm, length range 5–15μm) were purchased from Shenzhen Nanotech Port Co., Ltd., China (trade name of product is L-MWNTs-10). As-received MWCNTs were purified and mildly oxidized by boiling 30 wt% nitric acid for 1 h with subsequent settling at room temperature for 20 h. This procedure was carried out in order to remove rest of amorphous carbon and impurities that might be poisonous for metallocene catalysts and to increase content of carboxylic and hydroxyl groups on MWCNTs. The acid-treated MWCNTs were filtered and washed repeatedly with deionized water, dried in vacuum at 400°C for 5 h, and then stored in argon atmosphere.

Diameter of SWCNTs is 1.4 nm, and length is more than 5 μm.

Synthesis of nanocomposites was done in bulk propylene, as described in [18-19]. This metallocene catalyst used in the synthesis of composites is highly active and isospecific in propylene polymerization, producing iPP of high molecular weight [20]. The process was conducted at 60°C and pressure 2.5 MPa in a steel 200 cm$^3$ reactor vessel, equipped with a high-speed stirrer (3000 rev./min). Nanocomposites were synthesized via the following routes: 1) powder of carbon nanofiller (GNP, nanotubes or $C_{60}$), previously evacuated at 200°C, was fed into reactor vessel, which was then filled with liquid propylene (100 ml), methylalumoxane and metallocene catalyst; 2) GNP or TRGO was prepared as suspension in toluene and sonicated for 10 min, then methylalumoxane was added and sonication continued for 10 more minutes.

Ultrasonic power was ~ 35W. Afterwards the suspension was fed into reactor vessel, filled with liquid propylene and catalyst was finally added.

Concentration of filler in composites was varied by changing polymerization time. Final product was unloaded from the reactor, washed successively by a mixture of ethyl alcohol and HCl (10% solution), ethyl alcohol and then dried in vacuum at 60°C until constant weight.

Test specimen were cut from films 100–300 mm thick, pressure molded at 190°C and pressure 10 MPa at cooling rate 16 K·min$^{-1}$.

The SANS measurements were performed using the YuMO spectrometer at the IBR2 reactor in Dubna, Russian Federation [23].

The neutron wavelength is $\lambda$= 0.7–6Å, the neutron flux on a sample was about $10^7$ n/(s×cm$^2$) [24], diameter of neutron beam on the sample was 14 mm. The solid film-like specimens with different nanofillers iPP/GNP, iPP/nanographite, iPP/SWCNT, iPP/MWCNT, iPP/MWCNT/GNP, iPP/fullerene were fixed in the holder. The thickness values for the specimens were normalized to thickness of iPP film 368 μm. Holder was put into thermo box.

In the SANS measurements with YuMO, we recorded counts versus time of flight from the 16 rings of two detectors. Recalculation and normalization count using the gauge standard of the known cross section *vs* time of flight to the differential scattering cross section $d\Sigma/d\Omega(Q)$ and normalization on the sample thickness was realized by program SAS[25]. The preliminarily evaluated scattering length density (SLD) for the samples is in the range $3.84 \times 10^{10} - 5 \times 10^{10}$ cm$^{-1}$ [4].

## 3. Results and analysis of SANS spectra

To analyze the experimental small-angle scattering curves, we used a number of the following procedures of the ATSAS 2.4 software package [11].

Preliminary processing of the initial scattering curves and registration of scattering by the reference sample were performed using the PRIMUS procedure of ATSAS [11]. In this work, as the reference scattering, which was subtracted from the experimental curve of small-angle scattering of samples *I(Q)*, scattering from a sample of matrix polymer (iPP) was used. Thus, after taking into account reference scattering, the experimental small-angle scattering curves are characterized by scattering from only heterogeneous regions ("scattering particles") in the system having a scattering length different from the scattering length of the polymer matrix.

To calculate regularized scattering curves $I_{reg}(Q)$, optimized over the entire range of scattering angles, the particle distribution function, the integral values of the inertia radii of the particles of the scattering phase and the particle size distribution, we used the GNOM procedure of ATSAS based on the regularization method according to Tikhonov [12].

To determine the shape and spatial structure of scattering particles, we used an approach based on the use of well-founded algorithms for reconstructing the shape of scattering particles from small-angle scattering data, implemented in the DAMMIN and DAMMIF procedures of ATSAS, the algorithm of which uses the Monte Carlo method with the annealing procedure for reconstructing the shape of scattering particles in the framework of the model of "virtual" (dummy) atoms. The structures recovered in individual runs were averaged using the DAMAVER and SUPCOMB procedures [26].

### 3.1. SANS data for nanographite and GNP samples

The experimental values of the intensity $I(Q)$ of small-angle neutron scattering and the regularized small-angle X-ray scattering curves $I_{reg}(Q)$ calculated in accordance with the GNOM procedure for samples nanographite 1.8, 3.6 wt% and GNP 0.7, 1.8 wt%; (excluding scattering from iPP as a reference sample) are shown in Fig. 1.

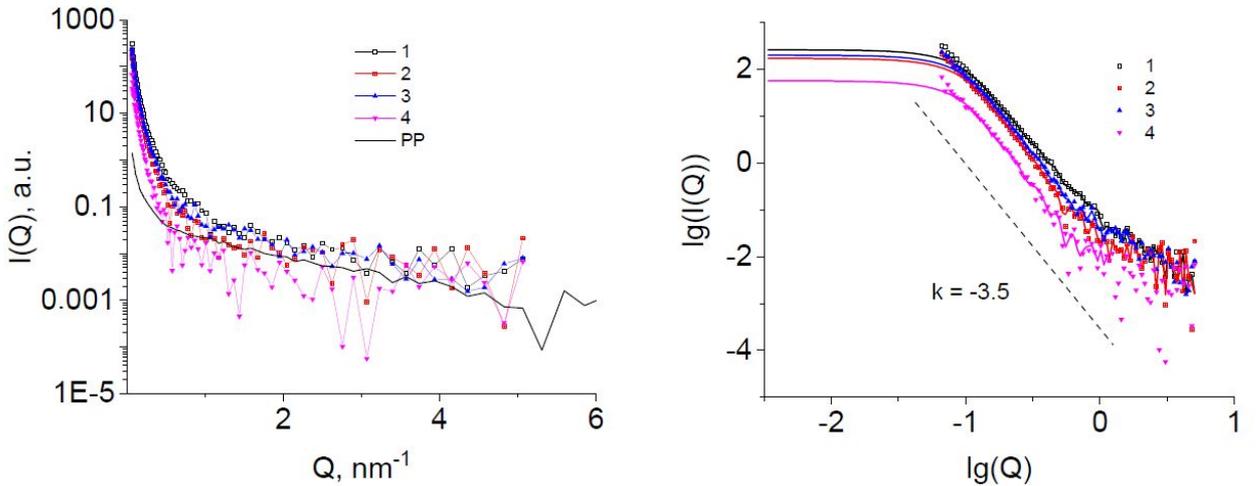

Fig. 1. The experimental SANS intensity $I(Q)$ of the samples in the coordinates $I$–$Q$ (left) and $\log(I)$–$\log(Q)$ (right). 1 – nanographite 3.6 wt%; 2 – nanographite 1.5 wt%; 3 – GNP 1.8 wt%; 4 – GNP 0.7 wt%. The solid lines correspond to the regularized $I_{reg}(Q)$ curves. The scattering curve of the matrix polymer iPP is shown for comparison. The value $k$ characterizes the slope of the linear sections of the scattering curves in the $\log(I)$ – $\log(Q)$ coordinates.

From Fig. 1. we conclude that scattering particles have an almost identical spatial structure and their scattering pattern corresponds to scattering by a physical fractal object with dimension $d_s = 6 - |k| = 2.5$, corresponding to a surface fractal. We reveal the dense compact aggregated particles with a rugged surface [13]. The upper size range of these physical fractals exceeds the spatial resolution of the small angle neutron scattering method $L_{max} = 2\pi/Q_{min} = 94$ nm, implemented in the experiments of this work ($Q_{min} = 0.0665$ nm$^{-1}$).

As there are no interference effects and the curves are of diffuse nature, small-angle scattering by a dilute or polydisperse system of particles can be interpreted with minimal detail of the scattering system, [4] either in the scattering approximation from a polydisperse system of particles with a known form factor (balls, prisms, cylinders, volume ellipsoids of revolution, *etc.*), or in the scattering approximation from identical particles of unknown shape and spatial structure. In the first case, the resulting interpretation of the small-angle scattering data is the reconstructed particle size distribution function, and in the second case, the determination of the shape and size of the particles. Note, that the possibility of restoring the low-resolution structure of polydisperse and polymorphic nano-objects having more than one hierarchical level of structural organization (particles - particle aggregates) from small-angle scattering data using the ATSAS software package was demonstrated earlier in [5,14].

Since the GNP and nanographite samples used in the work contain particles in the form of plates, the regularized scattering curves from these samples were primarily used to calculate the distribution function of particle thicknesses under the assumption of a polydisperse system of plate-shaped particles with thickness $T$ (the distance distribution function of thickness, assuming a polydisperse system of flat particles). The calculation results are shown in Fig. 2.

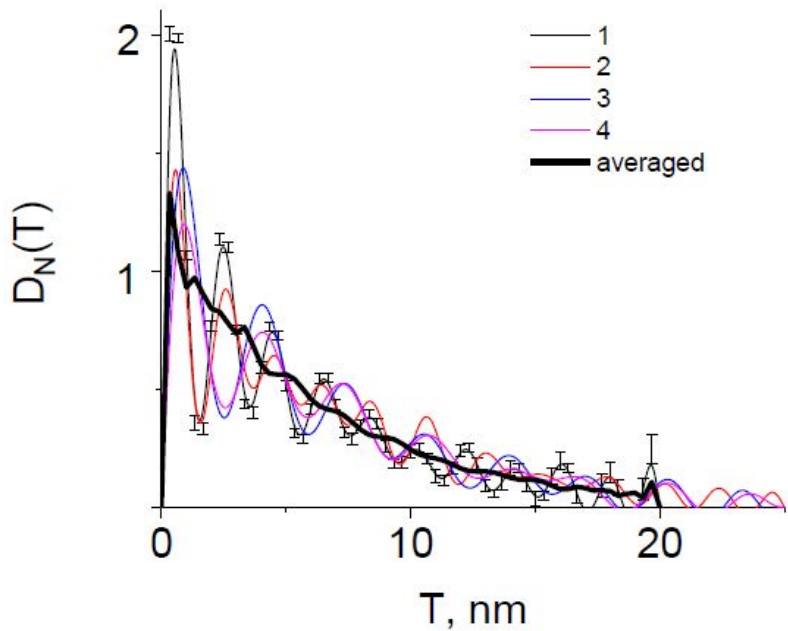

Fig. 2. Normalized distribution functions of particle thicknesses $D_N$ (thin lines) under the assumption of a polydisperse system of plate-shaped particles with a thickness $T$ calculated from the scattering curves for GNP and nanographite particles. The thick line is the smoothed curve, with adjacent averaging, weighted average value.

Similar to the scattering curves, the distribution functions for the GNP and nanographite samples turned out to be identical to each other. The system of scattering particles in the GNP and nanographite samples is characterized by high polydispersity. The particle size distribution contains wafers with a thickness of 1 to 20 nm, whereas the thickness of the initial plates are 47.3 nm and 1 nm for nanographite and GNP samples, respectively.

The radius of gyration $R_t = T^2/12$ of the particle thickness determined from the slope of the linear part of the $ln(Q^2I(Q))-Q^2$ plot (the Guinier plot) in the reciprocal space and that calculated by the indirect transform method [4] applied to the whole experimental scattering curve while using *GNOM* procedure, were close to each other and equal, on average, to 5.5 nm.

### 3.2. SANS data for the fullerene sample

The experimental values of the intensity *I(Q)* of small-angle neutron scattering and the regularized small-angle X-ray scattering curves $I_{reg}(Q)$ of the sample with 16.5wt% fullerene calculated with the GNOM procedure (excluding the scattering from iPP as a reference sample) are shown in Fig. 3 in the *I – Q* and log(*I*) – log(*Q*) coordinates.

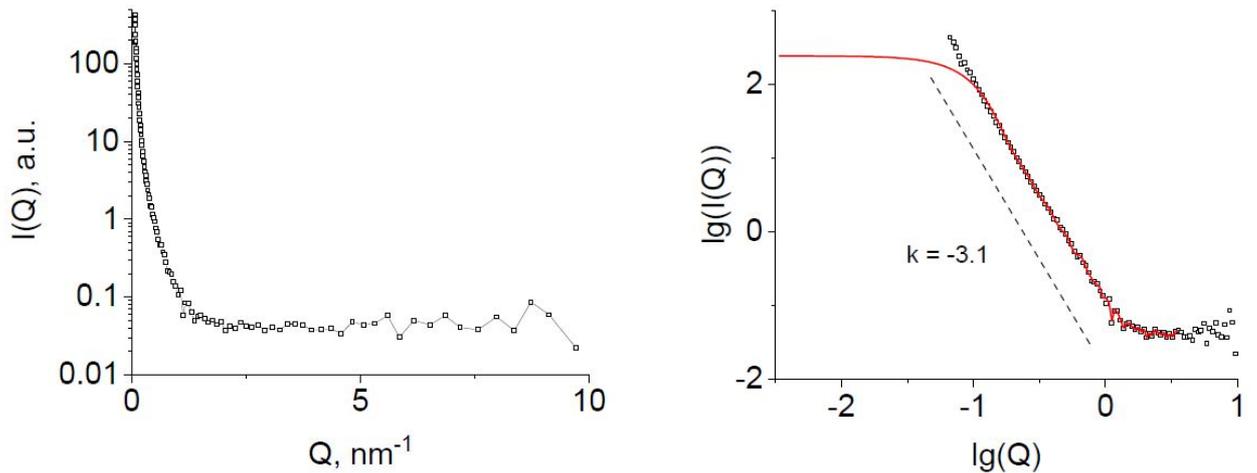

Fig. 3. The experimental SANS intensity *I(Q)* of the sample fullerene 16.5 wt%.

From Fig. 3 (right), we conclude that the scattering by fullerene particles of 16.5wt% concentration corresponds to scattering by a physical fractal object with dimension $d_s$ = 2.9, this is a surface fractal, where dense compact particles have a strongly rugged surface [13]). The upper size range of this physical fractal significantly exceeds the spatial resolution of the small-angle neutron scattering method $L_{max} = 2\pi/Q_{min}$ = 94 nm, realized in the experiments of this work.

The regularized scattering curve from the fullerene sample 16.5wt% was used to calculate the particle size distribution function assuming a polydisperse system of particles of a spherical shape with a radius *R* in the form of the volume distribution function of hard spheres. The calculation results are shown in Fig. 4.

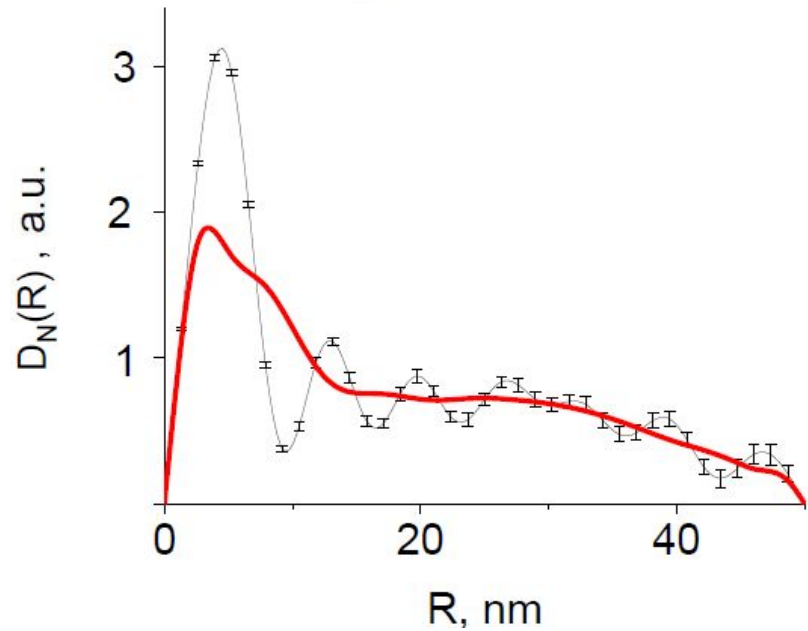

Fig. 4. Volume size particle distribution functions $D_V(R)$ for the fullerene 16.5 wt% scattering curve calculated under the assumption of a polydisperse system of spherical particles with radius $R$. The thin line denotes the calculated curve; and the thick one is the smoothed curve with adjacent averaging, weighted average value.

The radius of gyration $R_g$ of the particle determined from the slope of the linear part of the $ln(I(Q))$-$Q^2$ plot (the Guinier plot) in the reciprocal space and that calculated by the indirect transform method [4] applied to the whole experimental scattering curve while using the GNOM procedure were close to each other and equal, on the average, 29 nm.

The system of scattering particles in the fullerene 16.5wt% sample is characterized by high polydispersity. The particle size distribution contains spherical formations with a radius of 1 to 40 nm. The particles are aggregated. Note, that diameter of fullerene $C_{60}$ is 0.7 nm.

### 3.3. SANS data for MWCNT sample

The experimental values of the intensity $I(Q)$ of SANS and the regularized small-angle X-ray scattering curves $I_{reg}(Q)$ of the sample MWCNT 3.5wt% calculated with the GNOM procedure (excluding scattering from iPP as a reference sample) are shown in Fig. 5.

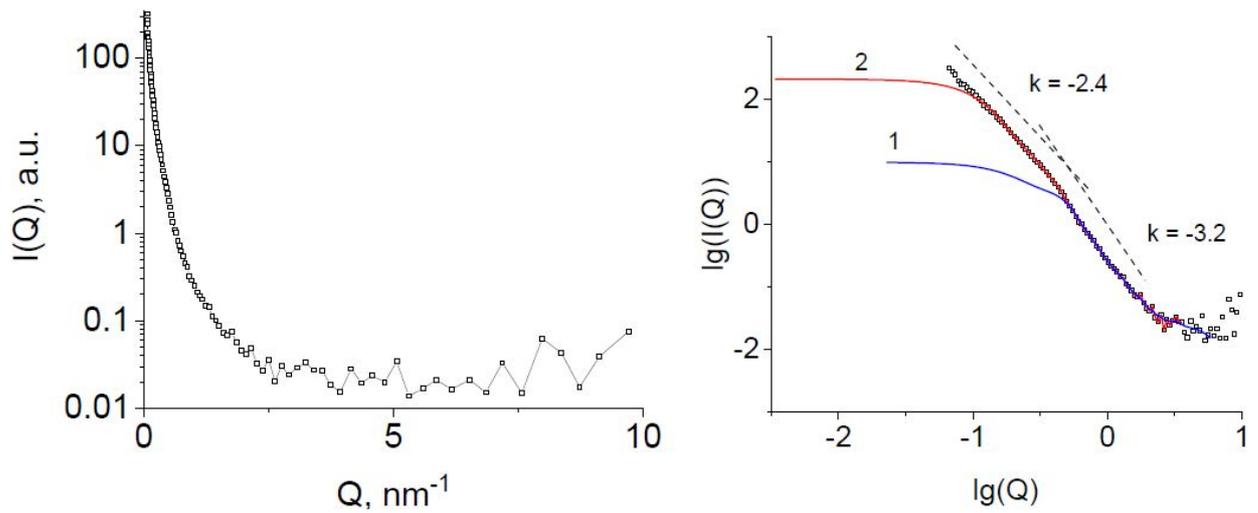

Fig. 5. The experimental SANS intensity *I(Q)* of the sample MWCNT 3.5 wt% in the coordinates *I–Q* (left) and log(*I*)–log(*Q*) (right). The label "1" denotes scattering from the MWCNT particles; "2" denotes scattering from aggregates of the MWCNT particles.

According to the procedure [5, 14], the MWCNT 3.5 wt% scattering curve (Fig. 5.) was divided into two regularized components, which relate to scattering by MWCNT particles (curve "1") and scattering from MWCNT aggregates (curve "2").

The relative content of aggregated and non-aggregated components in the system was estimated from the integrated intensity of the scattering components in the coordinates of the scattering intensity ($IQ^2$)–wave vector ($Q$). The calculated volume fractions of MWCNTs in the particle and aggregate forms were found to be equal to each other (~ 0.5). Both components of the SANS curve were analyzed independently.

The scattering curves "1" and "2" have been used to calculate the distance distribution function of the cross-section, assuming a polydisperse system of rod-like particles, as cylinders with radius *R* (Fig. 6).

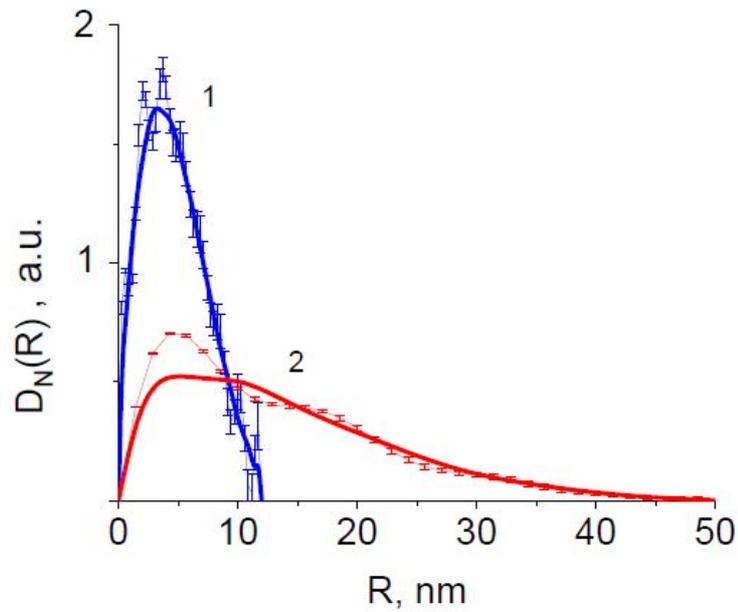

Fig. 6. The distribution function $D_N(R)$ of the MWCNT 3.5 wt% particle cross sections under the assumption of a polydisperse system of cylindrical particles with a radius $R$, calculated for scattering on the particles "1" and on the aggregates of particles "2". The thin line denotes the calculated curve; and the thick one is the smoothed curve with adjacent averaging, weighted average value.

The gyration radius $R_c$ of the particle cross-section determined from the slope of the linear part of the Guinier plot $\ln(qI(q))$-$q^2$ in the reciprocal space and calculated by the indirect transform method [5] applied to the whole experimental scattering curve while using the GNOM procedure, were close to each other and equal, on the average, to 3.8 and 11.4 nm for particles and aggregates, respectively, whereas the initial MWCNT diameter is 10 nm.

The results of reconstructing the shape and spatial structure of the system of particles and particle aggregates calculated from the regularized scattering curves of the MWCNT 3.5 wt% sample using the DAMMIN and DAMMIF procedures are presented in Fig. 7. Particle shape reconstuction was performed without any additional restrictions imposed on the expected symmetry and anisometry of the particles.

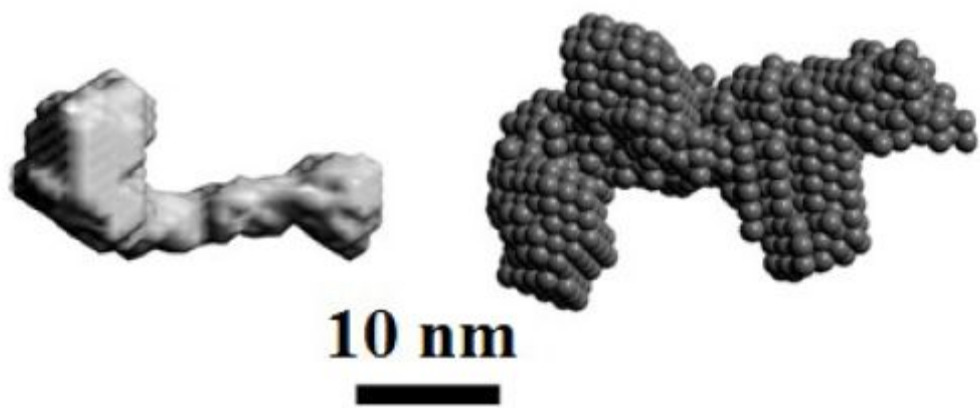

Fig. 7. The shape of the scattering particles calculated from the scattering curve of particles "1" (left) and aggregates of particles "2" (right). Visualization was performed with the model of bulk virtual ("dummy") atoms (right) and with the model of the surface accessible to solvent (left).

It is known [27] that small-angle scattering of a CNT system dispersed in a liquid medium or a polymer matrix is usually considered as scattering from some disordered two-level structure, in which the level of large characteristic sizes is referred to as CNT aggregates, and the level of small characteristic sizes, in turn, to straightened CNT fragments (analogue of the kinetic segment for the polymer chain), the persistent length of which is significantly less than the contour length of the CNT.

In this regard, the reconstructed form presented in Fig. 7 (left) for a scattering particle (a MWCNT fragment) in the form of an elongated cylinder with $D$ = 5–10 nm and a length $L$ = 30 nm is the expected result. In turn, according to results shown at Fig. 7 (right), the shape of scattering particles (aggregates) can most easily be interpreted as the "entanglements" of neighboring CNTs, analogous to similar "entanglements" of polymer macromolecules at concentrations higher than the crossover concentration. According toFig.5, the shape of scattering particles (aggregates) can most simply be interpreted as the "entanglements" of neighboring CNTs, by analogy with similar "entanglements" of polymer macromolecules at concentrations higher than the crossover concentration. In short, MWCNT system is similar to non-woven fabric structure.

### 3.4. SANS data for SWCNT samples

The experimental values of the SANS intensity $I(Q)$ of and the regularized SAXS curves $I_{reg}(Q)$ of the SWCNT 1.2, 2.6 and 8wt% samples(excluding scattering from iPP as a reference sample) calculated with the GNOM procedure are shown in Fig. 8.

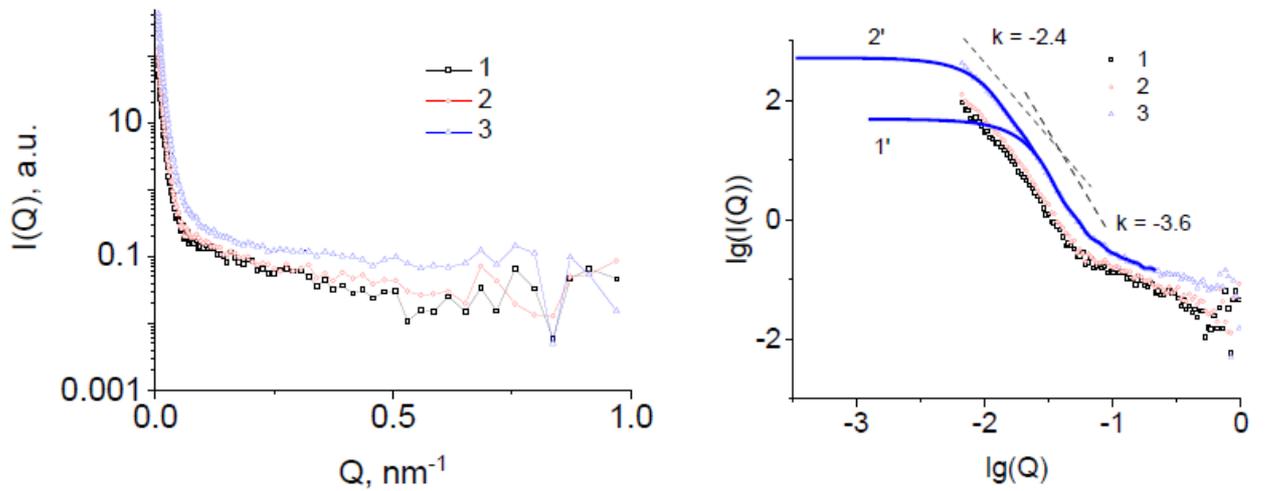

Fig. 8. The experimental SANS intensity $I(Q)$ for the samples SWCNT 1.2wt% (curves "1"), 2.6wt%(curves "2") and 8wt% (curves "3") in the coordinates $I - Q$ (left)and $\log(I)$–$\log(Q)$ (right). The labels "1'" and "2'" denote scattering from SWCNT particles and from aggregates of SWCNT particles respectively.

Fig. 8 shows, that scattering particles have an almost identical spatial structure.

According to the procedure [5, 14], we divided the scattering curves of the samples SWCNT 1.2 wt%, 2.6 wt% and 8 wt% (Fig. 8., right) into two regularized components that relate to scattering by SWCNT particles and scattering by SWCNT aggregates. To simplify the presentation, separation of the scattering curves into two components is shown in Fig. 8, right, only for sample SWCNT 8 wt%.

The relative content of aggregated and non-aggregated components in the system was estimated from the integrated intensity of the scattering components in the coordinates of the scattering intensity ($IQ^2$) – wave vector ($Q$). The calculated values of the volume fraction of SWCNT in the form of particles and aggregates for all samples SWCNT 1.2 wt%, 2.6 wt% and 8 wt% turned out to be equal to each other (~ 0.5). Both components of the scattering curve were independently analyzed.

The regularized scattering curves for SWCNT particles and particle aggregates were used to calculate the distribution function of particle cross sections under the assumption of a polydisperse system of cylindrical particles with a cylinder radius $R$ (the distance distribution function of the cross-section assuming a polydisperse system of rod-like particles). The calculation results are shown in Fig.9.

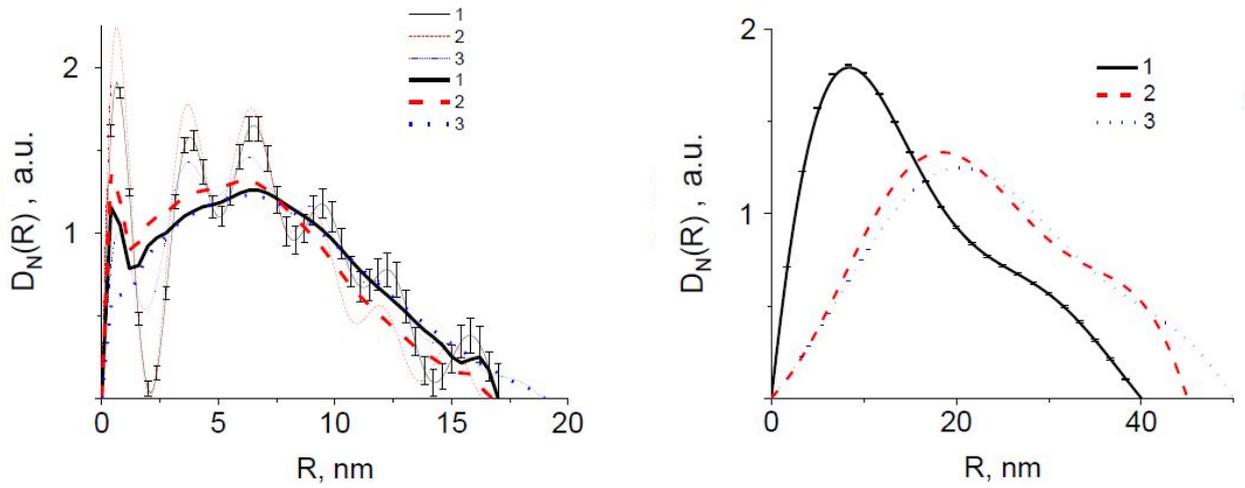

Fig. 9. The distribution function $D_N(R)$ of the cross sections of the particles of the samples SWCNT 1.2 wt%, (1), 2.6 wt% (2), and 8 wt% (3) under the assumption of a polydisperse system of cylindrical particles with a radius *R*, calculated for scattering on particles (left); and for aggregates of particles (right). The thin line denotes the calculated curve; and the thick one is the smoothed curve with adjacent averaging, weighted average value.

The gyration radius $R_c$ of SWCNT particle cross-section determined from the slope of the linear part of the Guinier plot $\ln(qI(q)) - q^2$ in the reciprocal space and calculated by the indirect transform method [4] applied to the whole experimental scattering curve using the GNOM procedure were equal to 5.6, 5.1, 5.8 nm for the samples SWCNT 1.2 wt%, SWCNT 2.6 wt%, SWCNT 8 wt%, respectively. Similarly, for the gyration radius $R_c$ of SWCNT particle aggregates cross-section, we have 12.3, 17.2, 18.2 nm for the same samples.

The results of reconstructing of shape and spatial structure of the SWCNT particle system and SWCNT particle aggregates calculated from the regularized scattering curves of the samples SWCNT 1.2 wt%, SWCNT 2.6 wt%, SWCNT 8 wt% using the DAMMIN and DAMMIF procedures are presented in Fig. 10. Particle shape reconstruction was performed without any additional restrictions imposed on the expected symmetry and anisometry of the particles.

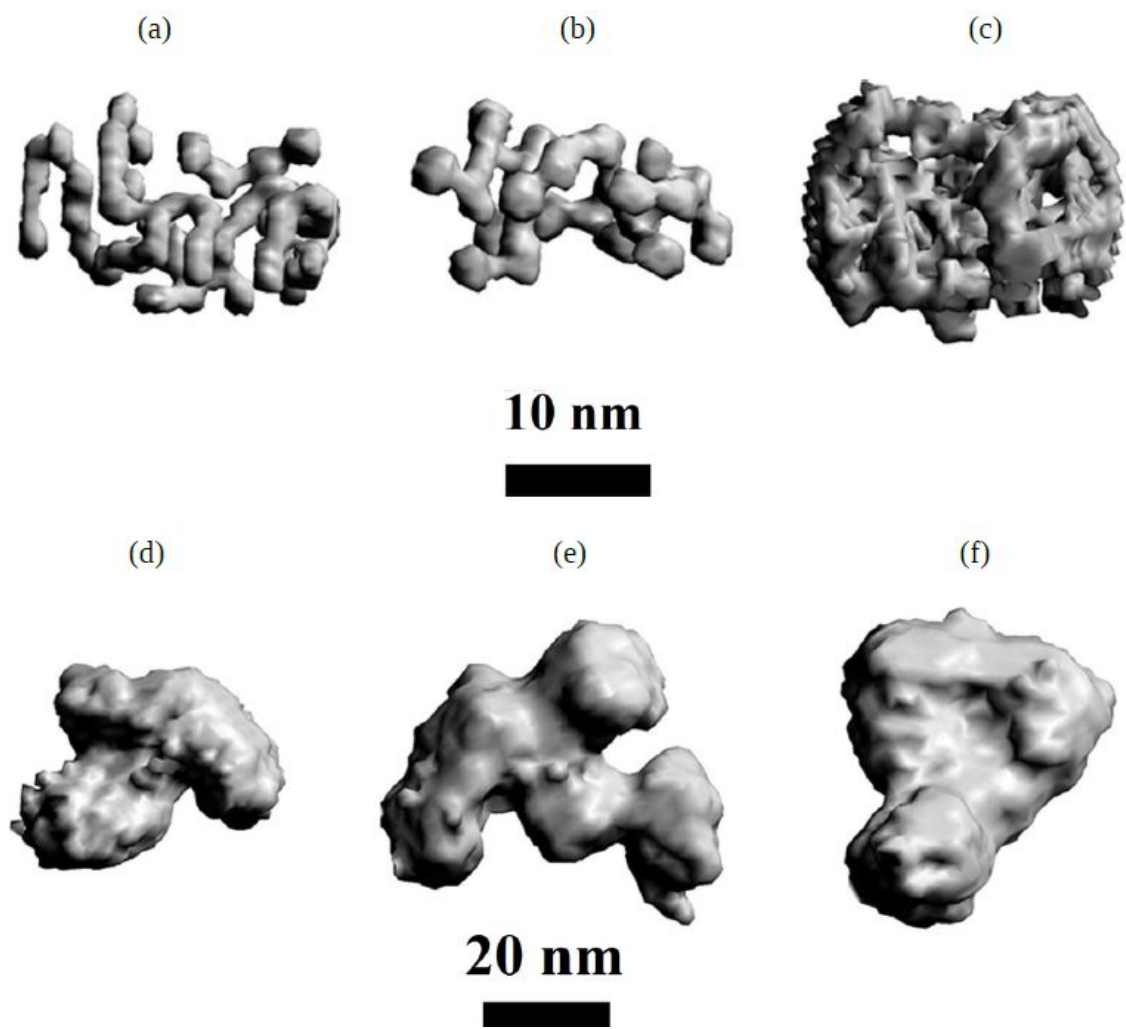

Fig. 10. The shape of the scattering particles calculated from the scattering curve of SWCNT particles (a–c) and SWCNT particle aggregates (d–e) of the samples SWCNT 1.2 wt% (a, d), SWCNT 2.6 wt% (b, d), SWCNT 8 wt% (c, f). Visualization with the model of bulk virtual ("dummy") atoms and the model of a surface accessible to a solvent

The reconstructed shape of the scattering particles reflects the quite expected compression of the spatial structure of the scattering system with increasing filler concentration. The SWCNT system can be described as a grid of individual tubes and tubes "stuck together" by the lateral surfaces and rather large knots.

### 3.5. SANS data for MWCNT/GNP samples

The experimental values of the SANS intensity $I(Q)$ of and the regularized SAXS curves $I_{reg}(Q)$ of the MWCNT/GNP 0.48 and 3 wt% samples (excluding scattering from iPP as a reference sample) calculated with the GNOM procedure are shown in Fig. 11.

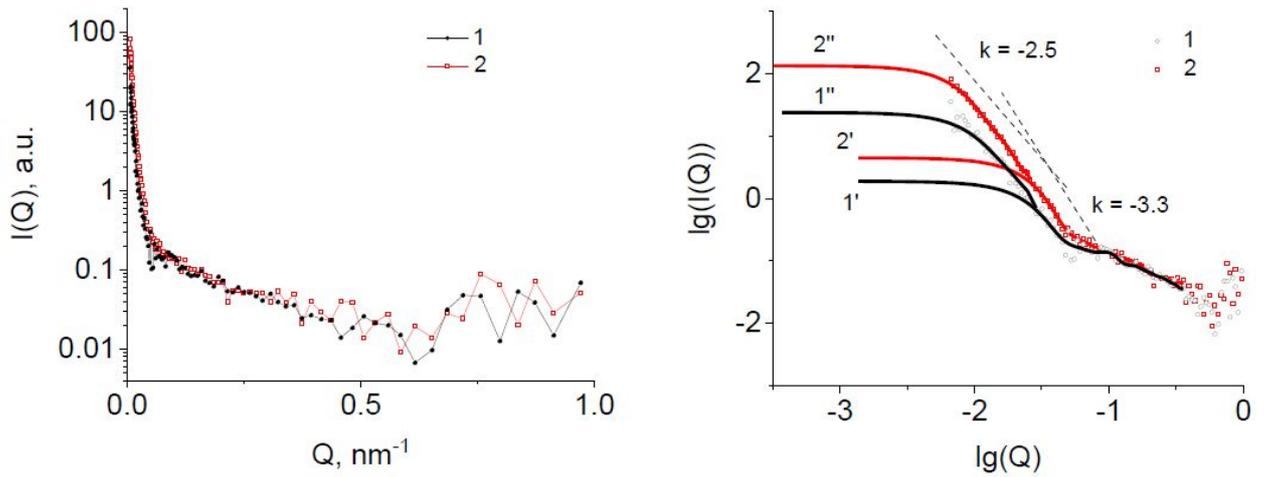

Fig. 11. The experimental SANS intensity *I(Q)* for samples MWCNT/GNP 0.48 wt% (1) and MWCNT/GNP 3 wt% (2) in the coordinates *I – Q* (left) and log (*I*) – log (*Q*) (right). The label "1'" denotes scattering from particles MWCNT/GNP 0.48 wt%; "1''" is scattering from particle aggregates MWCNT/GNP 0.48w%; "2'" is scattering from particles MWCNT/GNP 3wt%; "2''" is scattering from MWCNT/GNP 3wt% particle aggregates.

The scattering curves of the MWCNT/GNP 0.48 wt% and MWCNT/GNP 3 wt% samples (Fig. 11, right part) were divided into two regularized components [5, 14], which relate to scattering by particles and scattering by particle aggregates.

The relative content of aggregated and non-aggregated components in the system was estimated from the integrated intensity of the scattering components in the coordinates of the scattering intensity ($IQ^2$) – wave vector (*Q*). The calculated value of the volume fraction of particles and particle aggregates was 0.25/0.75 and 0.5/0.5 for samples MWCNT/GNP 0.48 wt% and MWCNT/GNP 3 wt% respectively. Both components of the scattering curve were independently analyzed for each of the samples.

Since the shape of the scattering particles was not known for "mixed" MWCNT/GNP samples, the regularized scattering curves for MWCNT/GNP particles and particle aggregates were used to calculate the most generalized characteristic: the particle volume distribution function under the assumption of a polydisperse system of spherical particles with a cylinder radius *R* (the volume distribution function of hard spheres). The calculation results are shown in Fig. 12.

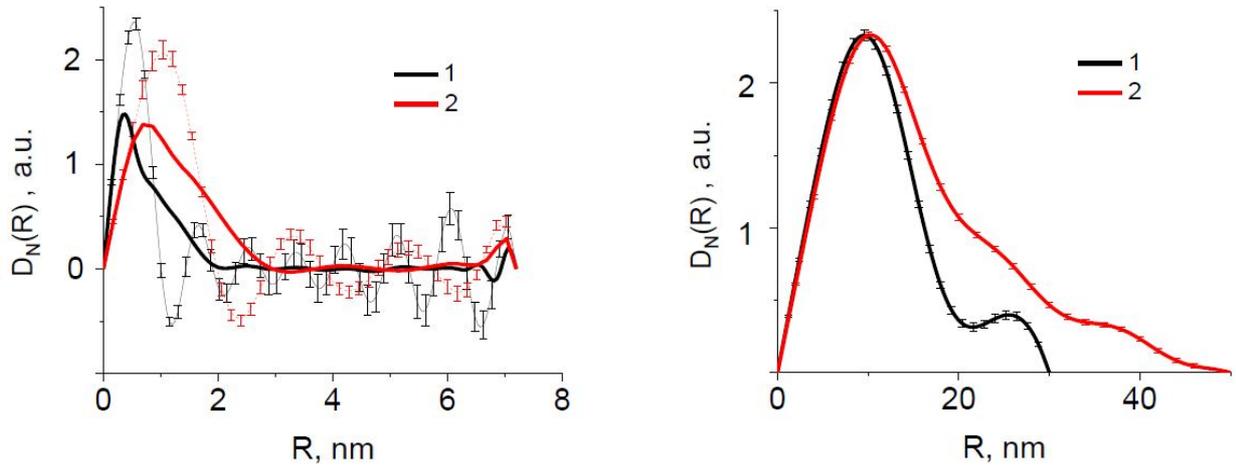

Fig. 12. Volume distribution function of the $D_N(R)$ particles of MWCNT/GNP 0.48 wt% samples (1) and MWCNT/GNP 3 wt% (2) under the assumption of a polydisperse system of particles of a spherical shape with a radius $R$, calculated for scattering: on particles (left) and on aggregates of particles (right). Thin and thick lines mean respectively calculated and smooth curves, adjacent averaging, weighted average.

The results of reconstructing the shape and spatial structure of the system of particles and particle aggregates calculated from the regularized scattering curves of the samples MWCNT/GNP 0.48 wt% and MWCNT/GNP 3 wt% using the DAMMIN and DAMMIF procedures are presented in Fig. 13. Particle shape reconstruction was performed without any additional restrictions imposed on the expected symmetry and anisometry of the particles.

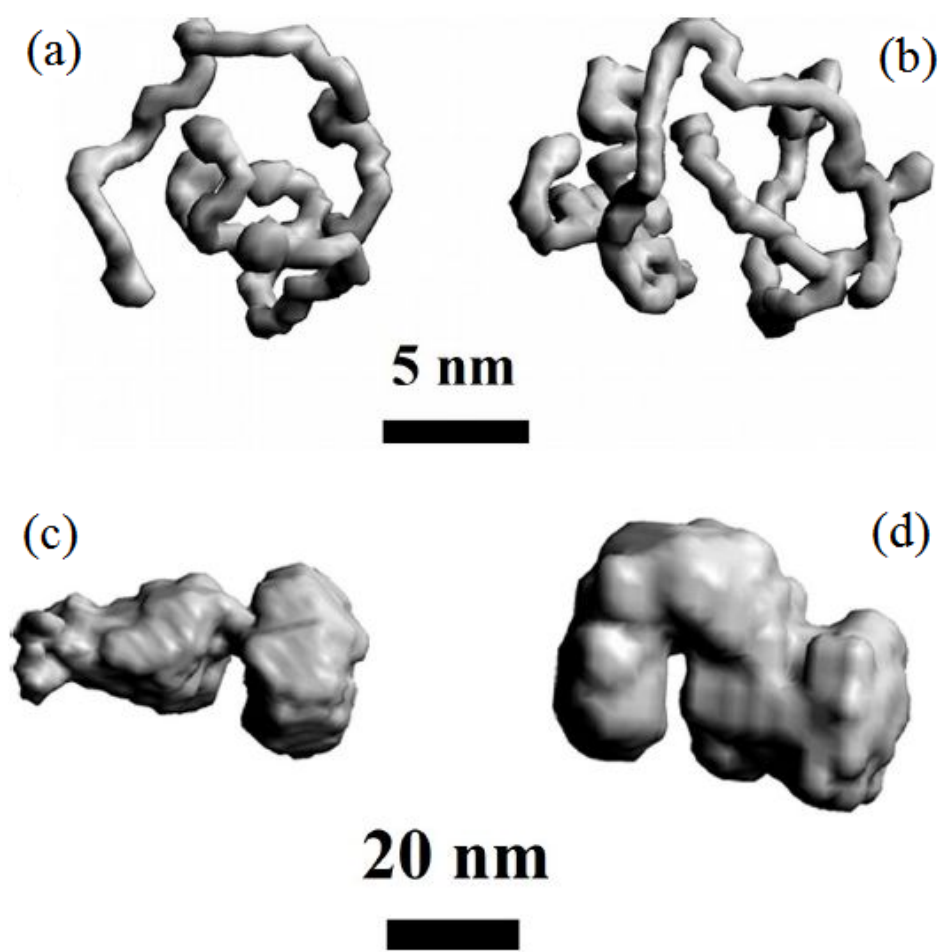

Fig. 13. The shape of scattering particles calculated from the scattering curve of particles (a, b) and particle aggregates (c, d) of MWCNT/GNP 0.48 wt% (a, c) and MWCNT/GNP 3 wt% (2) samples (b, d). Visualization with a model of bulk virtual ("dummy") atoms and a model of a surface accessible to a solvent.

The reconstructed shape of scattering particles in both samples refers to nanotubes; the shape of aggregates can be most likely ascribed to GNP. I.e., CNTs and GNP are separated in the iPP matrix. Moreover, the addition of GNP improves the dispersion of nanotubes.

4. **Conclusions**

The SANS method allowed us to characterize morphology of different carbon allotropes nanofillers in the volume of iPP in detail and to reconstruct their shape using numerical modeling excluding scattering from iPP. All nanofillers are found to be polydisperse, they form self-similar aggregates with high surface fractality. In iPP volume, CNTs twist in coils and knots, become more packaged, GNP and nanographite form new flat particles.

The system of scattering particles in the GNP and nanographite samples is characterized by high polydispersity. The particle size distribution contains wafers with a thickness of 1 to 20 nm, whereas the thickness of the initial plates is 47.3 nm and 1 nm for nanographite and GNP samples, respectively.

In the system of MWCNT the shape of scattering particles (aggregates) can most simply be interpreted as the "entanglements" of neighboring CNTs, by analogy with similar "entanglements" of polymer macromolecules at concentrations higher than the crossover concentration. In short, MWCNT system is similar to non-woven fabric structure.

The SWCNT system can be described as a grid of individual tubes and tubes "stuck together" by the lateral surfaces and rather large knots.

In case of binary filler (MWCNT+GNP) the shape of scattering particles refers to nanotubes, while the shape of aggregates can be most likely ascribed to GNP. I.e., CNTs and GNP are separated in the iPP matrix. Moreover, the addition of GNP improves the dispersion of nanotubes.

With modified geometry, nanofillers influence electron density distribution, electrical, optical, mechanical properties of functional basic polymers *etc.*, and this phenomenon requires further careful studies for practical applications

**Acknowledgements**

This work was supported by the Ministry of Science and Higher Education of the Russian Federation.

**References**

[1] Polymer Composites. V. 2.: Nanocomposites. Eds. T. Sabu, J. Kuruvilla, M. Sant Kumar, et al., Wiley-VCH Verlag GmbH & Co. KGaA.,2013.

[2] T. Hasan, V. Scardaci, P.H. Tan, F. Bonaccorso, A.G. Rozhin, Z. Sun and A.C. Ferrari, Nanotube and Graphene Polymer Composites for Photonics and Optoelectronics, in: O. Hayden, K. Nielsch (eds.), Molecular- and Nano-Tubes, Springer, 2011, pp. 279-354.DOI 10.1007/978-1-4419-9443-1_9.

[3] V.G. Shevchenko, S.V. Polschikov, P.M. Nedorezova, A.N. Klyamkina, A.N. Shchegolikhin, A.M. Aladyshev, V.E. Muradyan, In situ polymerized poly(propylene)/graphene nanoplatelets nanocomposites: dielectric and microwave properties, Polymer. 53 (2012) 5330–5335.

[4] L.A. Feigin, D.I. Svergun, Structure Analysis by Small-Angle X-ray and Neutron Scattering, Plenum Press, New York, 1987.

[5] A.N. Ozerin, T.S. Kurkin, L.A. Ozerina, V.Yu. Dolmatov, X-ray Diffraction Study of the Structure of Detonation Nanodiamonds, Crystallogr. Rep. 53 (1) (2008) 60–67. DOI:10.1134/S1063774508010070.

[6] J.S. Pedersen, Analysis of small-angle scattering data from colloids and polymer solutions: modeling and least-squares fitting, Advances in Colloid and Interface Sci. 70 (1997) 171–210.

[7] Y. Dror, W. Salalha, W. Pyckhout-Hintzen, A.L. Yarin, E. Zussman, Y. Cohen, From carbon nanotube dispersion to composite nanofibers, Progr. Colloid Polym Sci. 130 (2005) 64–69. DOI 10.1007/b107346.

[8] E.M. Milner, N.T. Skipper, C. A. Howard, M.S.P. Shaffer, D.J. Buckley, K.A. Rahnejat, P.L. Cullen, R.K. Heenan, P. Lindner, and R. Schweins, Structure and Morphology of Charged Graphene Platelets in Solution by Small-Angle Neutron Scattering, J. Am. Chem. Soc. 134 (2012) 8302–8305. dx.doi.org/10.1021/ja211869u.

[9] A. Manta, M. Gresil, C. Soutis, Predictive Model of Graphene Based Polymer Nanocomposites: Electrical Performance, Appl. Compos. Mater. 24 (2017) 281–300. DOI: 10.1007/s10443-016-9557-5.

[10] P.S. Singh, V.K. Aswal, Probing polymer nanocomposite morphology by small angle neutron scattering, Pramana – J. Phys. 71 (2008) 947–952.

[11] M.V. Petoukhov, D. Franke, A.V. Shkumatov, G. Tria, A.G. Kikhney, M. Gajda, C. Gorba, H.D.T. Mertens, P.V. Konarev, D.I. Svergun, New developments in the ATSAS program package for small-angle scattering data analysis, J. Appl. Cryst. 45 (2012) 342–350. doi:10.1107/S0021889812007662.

[12] A.N. Tikhonov, V.Ya. Arsenin, Solution of Ill-Posed Problems, Wiley, NY, 1977.

[13] J.E. Martin, A.J. Hurd, Scattering from Fractals, J. Appl. Cryst. 20 (1987) 61–78.

[14] E.V. Shtykova, Shape Determination of Polydisperse and Polymorphic Nanoobjects from Small-Angle X-Ray Scattering Data (Computer Simulation), Nanotechnologies in Russia. 10 (2015) 408–419. DOI: 10.1134/S1995078015030155.

[15] A.A. Arbuzov, V.E. Muradyan, B.P. Tarasov, Synthesis of graphene like materials by graphite oxide reduction, Russ. Chem. Bull.62 (2013)1962–1966.

[16] W.S. Hummers, R.E. Offeman, Preparation of Graphitic Oxide, Journal of the American Chemical Society 80 (1958) 1339-1339. http://dx.doi.org/10.1021/ja01539a017.

[17] V.E. Muradyan, V.S. Romanova, A.P. Moravsky, Z.N. Parnes, Yu.N. Novikov, A graphite oxide-based nickel catalyst for reductive dechlorination of polychlorinated aromatic hydrocarbons, Russian Chemical Bulletin. 49 (2000) 1017-1019. DOI: 10.1007/BF02494886.

[18] S.V. Polschikov, P.M. Nedorezova, A.N. Klyamkina, A.A. Kovalchuk, A.M. Aladyshev, A.N. Shchegolikhin, V.G. Shevchenko, V.E. Muradyan, Composite materials of graphene nanoplatelets and Polypropylene, prepared by in situ polymerization, J. Applied Polymer Sci. 127 (2013) 904-911.DOI: 10.1002/app.37837.

[19] P.M. Nedorezova, V.G. Shevchenko, A.N. Shchegolikhin, V.I. Tsvetkova, and Yu.M.

Korolev, Polymerizationally filled conducting polypropylene graphite composites prepared with highly efficient metallocene catalysts, Polym. Sci. Ser. A. 46 (2004) 242–249.

[20] W. Spaleck, F. Kuber, A. Winter, J. Rohrmann, B. Bochmann, M. Antberg, V. Dolle, E.F. Paulus, The Influence of Aromatic Substituents on the Polymerization Behavior of Bridged Zirconocene Catalysts, Organometallics. 13 (1994) 954–963.

[20] S.V. Polschikov, P.M. Nedorezova, O.M. Palaznik, A.N. Klyamkina, D.P. Shashkin, A.Ya. Gorenberg, V.G. Krasheninnikov, V.G. Shevchenko, A.A. Arbuzov, In situ polymerization of propylene with carbon nanoparticles. Effect of catalytic system and graphene type, Polymer Engineering & Science. 58 (2018) 1461-1470. DOI: 10.1002/pen.24644

[21] O.M. Palaznik, P.M. Nedorezovaa, S.V. Pol'shchikova, A.N. Klyamkina, V.G. Shevchenko, V.G. Krasheninnikov, T.V. Monakhova, and A.A. Arbuzov, Production by In Situ Polymerization and Properties of Composite Materials Based on Polypropylene and Hybrid Carbon Nanofillers, Polymer Science. Series B. 61 (2) (2019) 200–214.

[23] A.I. Kuklin, A.K. Islamov, V.I. Gordeliy, Scientific reviews: Two-detector system for small-angle neutron scattering instrument, Neutron News. 16(3) (2005) 16–18.

[24] A.I. Kuklin, A.D. Rogov, Yu.E. Gorshkova, P.K. Utrobin, Yu.S. Kovalev, A.V. Rogachev, O.I. Ivankov, S.A. Kutuzov, D.V. Soloviov, and V.I. Gordeliy, Analysis of neutron spectra and fluxes obtained with cold and thermal moderators at IBR-2 reactor: Experimental and computer-modeling studies, Physics of Particles and Nuclei Letters. 8.2 (2011) 119–128.

[25] A.G. Soloviev, T.M. Solovjeva, O.I. Ivankov, D.V. Soloviov, A.V. Rogachev, and A.I. Kuklin, SAS program for two-detector system: seamless curve from both detectors, Journal of Physics: Conference Series. 848 (2017) 012020-1-012020-7.
doi :10.1088/1742-6596/848/1/012020.

[26] M.B. Kozin and D.I. Svergun, Automated matching of high- and low-resolution structural models, J. Appl. Crystallogr. 34 (2001) 33–41.

[27] D.W. Schaefer, R.S. Justice, How Nano Are Nanocomposites?, Macromolecules. 40 (2007) 8501–8517.